\documentclass[aps, prb,twocolumn,superscriptaddress, nofootinbib]{revtex4-1}
\usepackage{graphicx}
\usepackage{dcolumn}
\usepackage{dcolumn}
\usepackage{textgreek}
\usepackage{textcomp}
\usepackage{adjustbox}
\usepackage{multirow}

\def\ECA{EuCd$_2$As$_2$}
\def\BCA{BaCd$_2$As$_2$}
\def\EBCA{Eu$_{1-x}$Ba$_x$Cd$_2$As$_2$}

\def\EnBCA{Eu$_{0.9}$Ba$_{0.1}$Cd$_2$As$_2$}

\def\EfBCA{Eu$_{0.5}$Ba$_{0.5}$Cd$_2$As$_2$}

\def\Tc{$T_c$}
\def\Tn{$T_N$}
\def\a{\textit{a}}

\def\c{\textit{c}}

\def\degrees{$^{\circ}$}

\def\Ptmo{$P\overline{3}m1$}

\begin{document}

\onecolumngrid
Notice of Copyright This manuscript has been authored by UT-Battelle, LLC under Contract No. DE-AC05-00OR22725 with the U.S. Department of Energy. The United States Government retains and the publisher, by accepting the article for publication, acknowledges that the United States Government retains a non-exclusive, paid-up, irrevocable, world-wide license to publish or reproduce the published form of this manuscript, or allow others to do so, for United States Government purposes. The Department of Energy will provide public access to these results of federally sponsored research in accordance with the DOE Public Access Plan (http://energy.gov/downloads/doe-public-access-plan).
\twocolumngrid

\clearpage

\preprint{APS/123-QED}

\title{Evidence of Ba substitution induced spin-canting in the magnetic Weyl semimetal EuCd$_2$As$_2$}


\author{L.D. Sanjeewa}
\thanks{These authors contributed equally}
\affiliation{Materials Science and Technology Division, Oak Ridge National Laboratory, Oak Ridge, TN 37831}
\author{J. Xing}
\thanks{These authors contributed equally}
\affiliation{Materials Science and Technology Division, Oak Ridge National Laboratory, Oak Ridge, TN 37831}
\author{K.M. Taddei}
\thanks{These authors contributed equally}
\affiliation{Neutron Scattering Division, Oak Ridge National Laboratory, Oak Ridge, TN 37831}
\email[corresponding author ]{taddeikm@ornl.gov}
\author{D. Parker}
\affiliation{Materials Science and Technology Division, Oak Ridge National Laboratory, Oak Ridge, TN 37831}
\author{R. Custelcean}
\affiliation{Chemical Sciences Division, Oak Ridge National Laboratory, Oak Ridge, TN 37831}
\author{C. dela Cruz}
\affiliation{Neutron Scattering Division, Oak Ridge National Laboratory, Oak Ridge, TN 37831}
\author{A.S. Sefat}
\affiliation{Materials Science and Technology Division, Oak Ridge National Laboratory, Oak Ridge, TN 37831}

\date{\today}

\begin{abstract}
Recently \ECA\ was predicted to be a magnetic Weyl semi-metal with a lone pair of Weyl nodes generated by A-type antiferromagnetism and protected by a rotational symmetry. However, it was soon discovered that the actual magnetic structure broke the rotational symmetry and internal pressure was later suggested as a route to stabilize the desired magnetic state. In this work we test this prediction by synthesizing a series of \EBCA\ single crystals and studying their structural, magnetic and transport properties via both experimental techniques and first-principles calculations.  We find that small concentrations of Ba ($\sim 3-10\% $) lead to a small out-of-plane canting of the Eu moment. However, for higher concentrations this effect is suppressed and a nearly in-plane model is recovered. Studying the transport properties we find that all compositions show evidence of an Anomalous Hall Effect dominated by the intrinsic mechanism as well as large negative magnetoresistances in the longitudinal channel. A non-monotonic evolution of the transport properties is seen across the series which correlates to the proposed canting suggesting canting may enhance the topological effects. Careful density functional theory calculations using an all-electron approach revise prior predictions finding a purely ferromagnetic ground state with in-plane moments for both the \ECA\ and \EfBCA\ compounds - corroborating our experimental findings. This work suggests that Ba substitution can tune the magnetic properties in unexpected ways which correlate to changes in measures of topological properties, encouraging future work to locate the ideal Ba concentration for Eu moment canting.  
 
\end{abstract}


\maketitle


\section{\label{sec:intro}Introduction}

Topological materials have garnered significant interest in the condensed matter community for their potential to unlock theoretically proposed computing paradigms via the realization of physics long constrained to the context of high energy physics \cite{Wen2017}. This broad novelty arises as topological materials eschew Landau classification and instead host electronic states engendered by topological invariants rather than broken symmetries giving rise to novel quasiparticles such as Dirac, Weyl and  Majorana Fermions or even exotic long range entangled states \cite{Wen2017,Bansil2016,Burkov2016}. 

Of the former group of symmetry protected topological sates, Weyl semi-metals (WSM) have received significant interest due to their potential applications in quantum devices ranging from photovoltaics to valleytronics to quantum computing \cite{Osterhoudt2019, Chan2016,Chan2017, Parameswaran2014,Kharzeev2019,Jia2016}. However, while WSM have great promise and numerous candidate materials have been discovered, there is some inherent difficulty in both finding unambiguous signatures of the Weyl physics and harnessing those properties for applications. One reason for this difficulty arises due to the presence of non-topological Fermi surfaces in many WSM in addition to the Weyl points which lead to additional quasiparticles that can obscure the signatures of the Weyl physics \cite{Halasz2012,Armitage2018,Ramshaw2018}. Additionally - and more detrimentally - many WSM have multiple pairs of Weyl points which can thwart obvious signatures of the desired physics \cite{Armitage2018, Wang2017,Yan2017}. This second issue is partially inherent to the physics - for the majority of WSM where inversion symmetry (IS) breaking is responsible for creating the Weyl points the minimal number of pairs of Weyl points is constrained to 4 and is often much higher such as 24 in the familiar TaAs compound \cite{Weng2015,Armitage2018}. The occurrence of numerous pairs of Weyl points leads to a corresponding complexity in their signatures which are then less easily identified. 

However, there is a second class of WSM in which time-reversal symmetry breaking gives rise to the Weyl physics \cite{Wang2010ti,Hughes2011}. In these magnetic Weyl semi-metals (MWSM) a minimum number of Weyl points - two - is possible \cite{Armitage2018}. If a material which realized this minimal case could be found it would allow for a more direct observation and potentially control of the Weyl physics \cite{Armitage2018}. Unfortunately, MWSM have proven more difficult to find than the IS breaking variety and of those discovered, many still manifest multiple pairs of Weyl nodes complicating their study \cite{Liu2019,Chang2018}. 

Recently, \ECA\ (with centrosymmetric space group symmetry \Ptmo ) was predicted to be a MWSM with the minimal number of Weyl nodes \cite{Hua2018}. Sorting through magnetic space group symmetries and their effects on the nodal Fermi surfaces, Hua \textit{et al.} found that if a magnetic order was realized in \ECA\ which preserved the $C_{3}$ rotation axis via A-type antiferromagnetism (AFM) (i.e. \textit{c}-polarized Eu moments ferromagnetically (FM) coupled in-plane and AFM coupled out-of-plane) \ECA\ would go from a Dirac semi-metal to a MWSM which realized only a single pair of Weyl nodes \cite{Hua2018}. 

In the work that followed a more complex story emerged. Initial characterizations based on transport and magnetization measurements suggested the desired magnetic order with \Tn\ $\sim 10$ K however, followup reports using resonant x-ray scattering and density functional theory (DFT) calculations suggested the wanted FM and AFM intra and inter-layer correlations but with in-plane Eu moments \cite{Schellenberg2011, Wang2016a,Krishna2018,Rahn2018,Soh2019}. Nonetheless, Hall Effect (HE) measurements suggested a significant and gate-controllable Anomalous Hall Effect (AHE) and studies using applied magnetic fields showed that the \textit{c}-polarized state could be achieved by applying a modest ($\sim 1.5$ T) field along the \textit{c} axis leading to sustained interest in \ECA \cite{Niu2019,Wang2016a, Rahn2018}. 

Yet more recently, a careful study using muon spin rotation to probe the magnetic fluctuations suggested a novel behavior above \Tn\ where intra-layer magnetic interactions stabilized long-lived fluctuations of the \textit{c}-polarized state which could generate the desired Weyl physics (identified via angle resolved photoemission spectroscopy) \cite{Ma2019}. Furthermore, additional DFT work proposed that weakening the inter-layer Eu coupling via internal pressure though Ba substitution could stabilize the \textit{c}-polarized FM state \cite{Wang2019}. In their calculations Wang \textit{et al.} suggest a Ba concentration of 50\%\ (if assumed to separate into ordered Ba only and Eu only layers) could dilate the \textit{c}-axis enough to achieve the needed reduced inter-layer coupling. Concurrently, a new synthesis study suggested another route to FM via the introduction of Eu vacancies \cite{Jo2020}. In this work it was suggested that the in-plane AFM could be tuned to FM with a \Tc\ of $\sim 30$ K via a small $\sim 1$\%\ Eu deficiency thus adding another avenue to potentially achieve the desired phase \cite{Jo2020}.

In this report, we contribute to this search by following up on Wang \textit{et al.}'s prediction, growing a series of Ba substituted \ECA\ (\EBCA\ with $x\leq 0.4$.) Using transport, magnetization, x-ray diffraction and DFT we find that while the Ba substitution drives the expected \textit{c}-axis dilation it has a quite different effect on the magnetism than initially predicted - leading to a small canting of the Eu moments out-of-plane but only for small $x\leq 0.3$ concentrations. Furthermore, transport measurements show that over the whole series the intrinsic contribution dominates the AHE and that the proposed canting seems to correlate to increased (and significant) negative magnetoresistances (nMR) in the longitudinal channels. Finally, revisiting the DFT calculations we find that the parent compound exhibits a small in-plane magnetic anisotorpy and is predicted to have the in-plane FM (as found in our experiments) as its ground state, though not owing to Eu vacancies. Turning to \EfBCA\ we find little evidence to motivate a layer separated state and that for either this assumed state or a random occupancy state the substitution of Eu with Ba actually stabilizes AFM. Nonetheless, our results show that Ba substitution can tune the magnetic structure - though in a more complicated fashion than previously proposed - and that that tuning correlates to promising changes in measurements expected to chart the topological properties.

\section{\label{sec:methods} Methods}

\subsection{\label{subsec:synthesis} Synthesis}

Single crystals of \ECA\ were grown using a flux method via a molten-salt media of KCl/NaCl (in a equimolar mixture.) The reactants, Eu-pieces, Cd-pieces (Alfa Aesar, 99.99\%) and As-pieces (Alfa Aesar, 99.99\%), were mixed with the salt flux in a He-purged dry box. In a typical reaction, a total of 1 g of Eu, Cd and As were used in a stoichiometric ratio of 1 : 2: 2 with 10 g of total flux of KCl/NaCl. The reaction mixture was sealed in an evacuated fused-silica ampule and then heated to 497 \degrees C at a rate of 20 \degrees C/hr and held for one day, followed by another heating to 597 \degrees C at a rate of 20 \degrees C/hr and held for one day. After that, reactions were heated to 847 \degrees C, 20 \degrees C/hr and held for 100 hrs. As the final step, reactions were cooled to 500 \degrees C at 1 \degrees C /hr, and then furnace cooled to room temperature. Black shiny plate crystals were recovered by washing the product with deionized water using the vacuum filtration method. Single crystals were physically examined and selected under an optical microscope equipped with a polarizing light attachment (see fig.~\ref{fig:one}.) For \EBCA\ the same method was employed with the appropriate stoichiometric amounts of Eu and Ba.

\subsection{\label{subsec:char} Characterization}

For single crystal x-ray diffraction (SXRD) studies, single crystals of \EBCA\ were sonicated in acetone to remove any surface impurities. The SXRD was performed by Bruker Quest D8 single-crystal X-ray diffractometer. The data were collected at room temperature utilizing Mo K$\alpha$ radiation, $\lambda$ = 0.71073 \AA. The crystal diffraction images were collected using $\phi$ and $\omega$ -scans. The diffractometer was equipped with an Incoatec I$\mu$S source using the APEXIII software suite for data setup, collection, and processing\cite{Bruker2015}. All the structures were resolved using intrinsic phasing and full-matrix least square methods with refinement on $F^2$. All of the structure refinements were done using the SHELXTL software suite \cite{Sheldrick2008}. All atoms were first refined with isotropic atomic displacement parameters which were later refined anisotropically.

Energy-dispersive spectroscopy analysis (EDS) was performed using a Hitachi S3400 scanning electron microscope equipped with an OXFORD EDX microprobe to confirm the elemental composition in all single crystal samples. Powder x-ray diffraction (PXRD) data were collected using a PANalytical X’Pert Pro MPD diffractometer with Cu K$\alpha 1$ radiation ($\lambda$ = 1.5418 \AA) at room temperature. PXRD pattern of \ECA\ and \EBCA\ were collected to identify the crystallographic plane of the flat surface of single crystals. Additionally, powder neutron diffraction patterns were collected for \EnBCA\ on HB-2A of Oak Ridge National Laboratory's High Flux Isotope Reactor to attempt magnetic structure solution, as a further check of composition and to test the feasibility of neutron diffraction in a non-isotopic sample \cite{Calder2018}. These data, however, proved difficult to analyze even qualitatively due to the large neutron absorption cross-sections of both Eu and Cd (see the supplemental materials) \cite{SM}.

Temperature and field-dependent magnetic measurements were carried out using a Quantum Design Magnetic Property Measurement System (MPMS). The measurements were carried out on single crystal specimens of \ECA\ and \EBCA\ samples by orienting the crystals such that the Eu-Eu-layers were aligned either parallel ($H\perp c$) or perpendicular ($H||c$) to the applied magnetic field. The temperature dependence of static susceptibility $[M/H(T)]$ was measured over a temperature range of 2–50 K for the applied fields of 50 Oe. Isothermal magnetization measurements were performed at 2 K for fields up to 60 kOe.

Resistivity and Hall effect measurements were performed in a Quantum Design Physical Properties Measurement System (PPMS) using the standard four point method. To ensure good results, the contact resistances were checked before each measurement and measurements were performed on multiple crystals of each composition. The crystal sizes were measured using a Leika microscope with a camera leading to an error of $\sim 0.1$ mm.      

\subsection{\label{subsec:DFT} First Principles Calculations}

In an attempt to make sense of the magnetic and related behaviors observed in this system, we performed first principles calculations using the all-electron plane-wave all-electron DFT code WIEN2K \cite{Blaha2018}. We began with the standard generalized gradient approximation (GGA) for structural refinements and moved to the GGA+U approach, with a U value of 5 eV applied to the Eu 4\textit{f} orbitals, as used in previous work \cite{Perdew1996,Wang2019}. 

For the structure we used the lattice parameters of 4.45 and 7.35 \AA\ for \textit{a} and \textit{c} respectively (following the convention of previous work) \cite{Wang2019, Schellenberg2011}.  We then relaxed the internal coordinates (with neither spin-orbit nor a `U') within the unit cell with the one Eu spin-polarized (with a moment of essentially 7 $\mu_B$), given that we are studying similar configurations for their magnetic characteristics. While we do not expect strong magnetoelastic coupling here given the generally localized nature of the magnetic Eu 4\textit{f} orbitals (unlike recent work on 3\textit{d}-based magnetic materials), it is best to take extra care with this computationally difficult system \cite{Sefat2016,Pokharel2018,Chen2019}. 


\section{\label{sec:res} Results \& Discussion}

\subsection{\label{crystal} Crystallographic Properties}

\begin{figure}
	\includegraphics[width=\columnwidth]{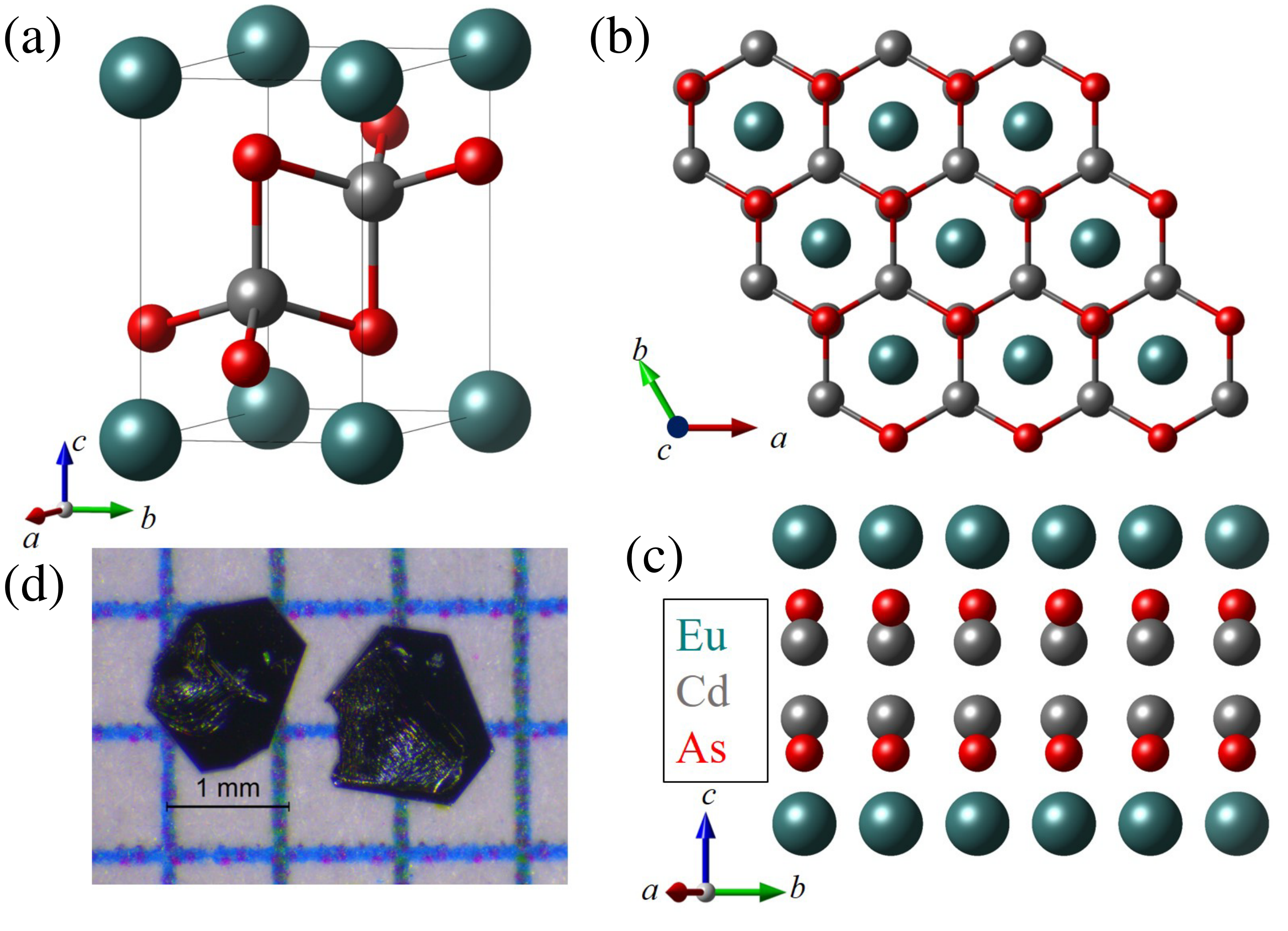}
	\caption{\label{fig:one}  Crystal lattice of \ECA\ viewed (a) parallel and (b) perpendicular to the Eu planes. (c) \ECA\ oriented to show the Eu-As-Cd-Cd-As-Eu stacking sequence. Eu, Cd and As atoms are shown in teal, gray and red respectively. (d) shows representative shape and size of \EBCA\ crystals.}	
\end{figure}

The nuclear crystal structure of \ECA\ is shown in Fig.~\ref{fig:one}. \ECA\ crystallizes in the centro-symmetric \Ptmo\ space group with Eu occupying the crystallographic $1a (0,0,0)$ Wyckoff position forming trigonal layers that alternate with layers of distorted edge-sharing CdAs$_4$ tetrahedra (where both Cd and As occupy the $2d (\frac{1}{3},\frac{2}{3},z)$ Wyckoff site.) The structure can be thought of as stacked triangular mono-atomic layers of order Eu-As-Cd-Cd-As-Eu. In each layer the inter-atomic spacing is equivalent to the in-plane lattice parameter \textit{a} with the inter-layer distances varying due to the free $z$ position of the Cd and As sites (Fig.~\ref{fig:one}(c)).  

\begin{figure}
	\includegraphics[width=\columnwidth]{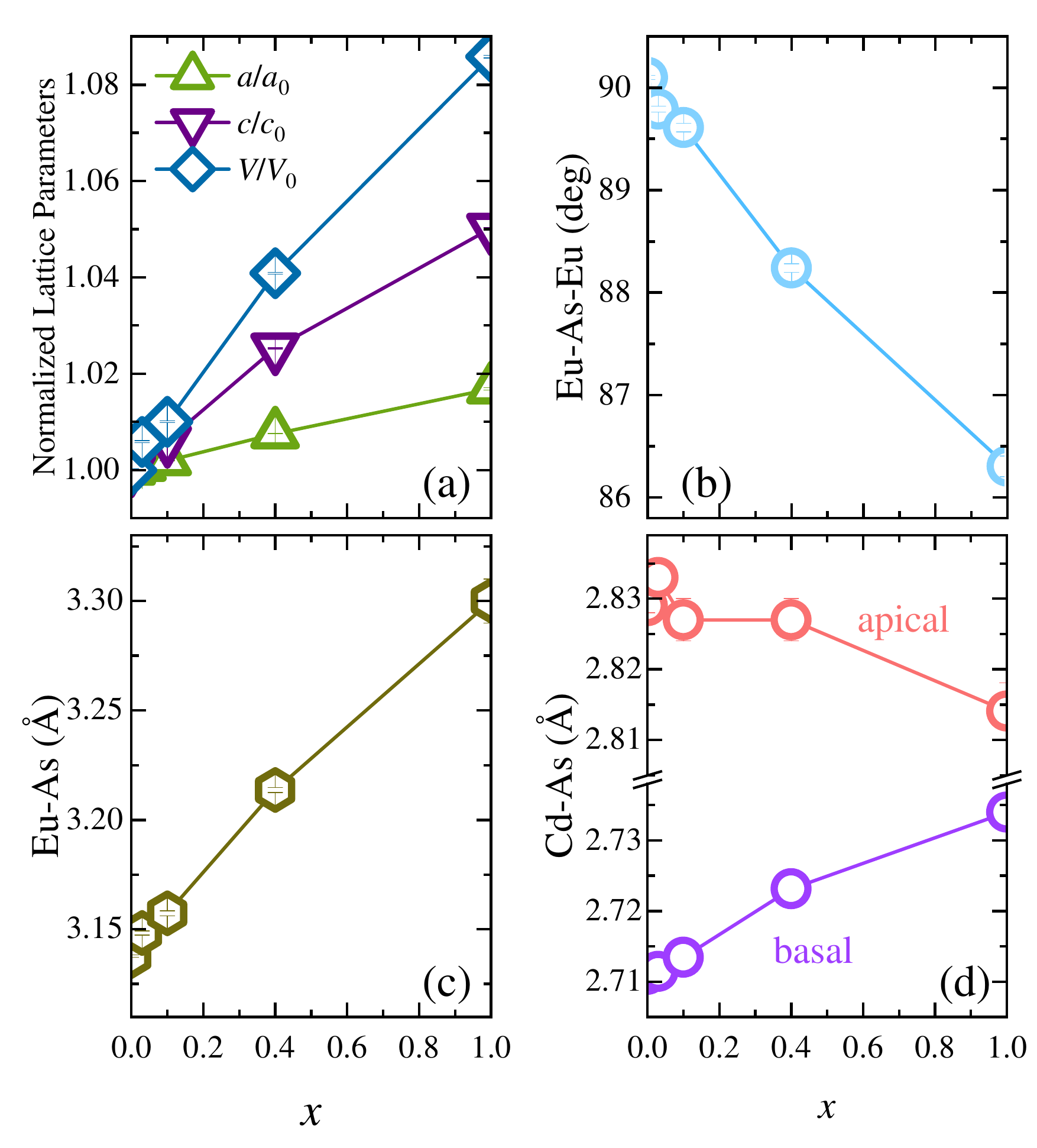}
	\caption{\label{fig:two}  Substitution dependence (in fractional Ba substitution) of the (a) normalized lattice parameters (normalized to the \ECA\ parent compound) (b) the Eu-As-Eu bond angle (c) the Eu-As bond length and (d) the basal and apical Cd-As bond lengths. Crystallographic parameters for the \BCA\ endmember were taken from ref.~\onlinecite{Klufers1984} }	
\end{figure}

In Fig.~\ref{fig:two} we report crystallographic parameters for \EBCA\ with x = 0, 0.03, 0.1 and 0.4  extracted from Rietveld refinements performed using SXRD data collected at room temperature (numerical values are reported in the SM) \cite{SM}. Starting with the parent compound, we find unit cell parameters similar to those reported previously with $a$ and $c$ of 4.4412  \AA\ and 7.3255 \AA\ respectively\cite{Artmann1996,Schellenberg2011}. For the internal parameters, the distorted CdAs$_4$ tetrahedra have one \lq apical\rq\ and 3 \lq basal\rq\ Cd-As bonds, the former of which is parallel to the \textit{c} axis. Our refinements find the basal bonds to be 4\%\ shorter at 2.7110(4) \AA\ relative to 2.829 (1) \AA\ for the apical bond. Considering the Eu local environment, there is only one Eu-As bond at 3.1377 \AA\ and one Eu-As-Eu bond angle, at 90.1(2)\degrees\ - the latter of which is consistent with in-plane FM interactions considering the Goodenough-Kanamori rules for exchange interactions these are, however, not always applicable to rare-earth super-exchange \cite{Geertsma1990}.

Upon substituting Ba on the Eu site, we observe qualitatively the expected expansion of the lattice suggested by Wang \textit{et al} in ref.~\onlinecite{Wang2019} (see Fig.~\ref{fig:two}(a).) The effect is anisotropic with the \a\ axis expanding less than the \c\ axis. Taking the 40\%\ sample we find a $\sim$ 0.5\%\ expansion of the \textit{a} axis and a $\sim$ 2.5\%\ expansion along \textit{c} with an overall increase in the unit cell volume of $\sim$ 4\% . We note that this volume expansion is similar to the 4\%\ expansion predicted by Wang \textit{et. al.} as needed to stabilize an out-of-plane magnetic order however, our parent compound has a somewhat smaller unit cell than their DFT calculations and so our raw lattice parameters in the 40\%\ are smaller than that predicted for the FM out-of-plane structure. Considering the lattice parameters separately our 40\%\ sample's \c\ axis is very near that predicted by ref.~\onlinecite{Wang2019} for \EfBCA . On the other hand, Wang \textit{et al.}'s predicted \a\ axis for the ideal \EfBCA\ sample is actually larger than that reported for \BCA\ which we think unlikely in the absence of a structural transition upon Ba substitution \cite{Klufers1984}.

Considering the bonding parameters, the Ba doping has little effect on the CdAs$_4$ tetrahedra, with the apical bond remaining essentially unchanged across the series and the three basal bonds expanding roughly commensurate with the change in the \a\ axis (Fig.~\ref{fig:two}(d).) On the other hand, the Eu-As bond continually dilates with doping and the Eu-As-Eu bond angle (Fig.~\ref{fig:two}(b)) is found to diverge from the nearly 90\degrees\ found in the parent compound, monotonically decreasing to smaller angles with Ba substitution. Presumably, these latter changes may be the more significant as the Eu-As-Eu bond angle describes the super-exchange pathway and so may affect the ground state magnetic structure. 

As a check of our nominal compositions, we performed EDS analysis which showed the compositions to agree with the nominal stoichiometry to within the certainty of the technique. As a second check, we also compared the obtained lattice parameters for all samples to those reported previously for \ECA\ and \BCA\ using Vegard's law \cite{Artmann1996,Schellenberg2011,Klufers1984,Vegard1921}. Doing so results in calculated Ba concentrations in reasonable agreement with the nominal stoichiometry and EDS analysis. As will become clear, it is important to note that we found no evidence of Eu vacancies in our parent compound in any of these tests. Furthermore, despite the predictions of ref.~\onlinecite{Wang2019}, we saw no evidence of Eu-Ba ordering in our structure solution work though we did not achieve the \EfBCA\ composition exactly, as we will discuss later, we believe the Eu-Ba ordered phase to be unlikely to stabilize using the current synthesis techniques.           

\subsection{\label{mag}Magnetic properties}

\begin{figure}
	\includegraphics[width=\columnwidth]{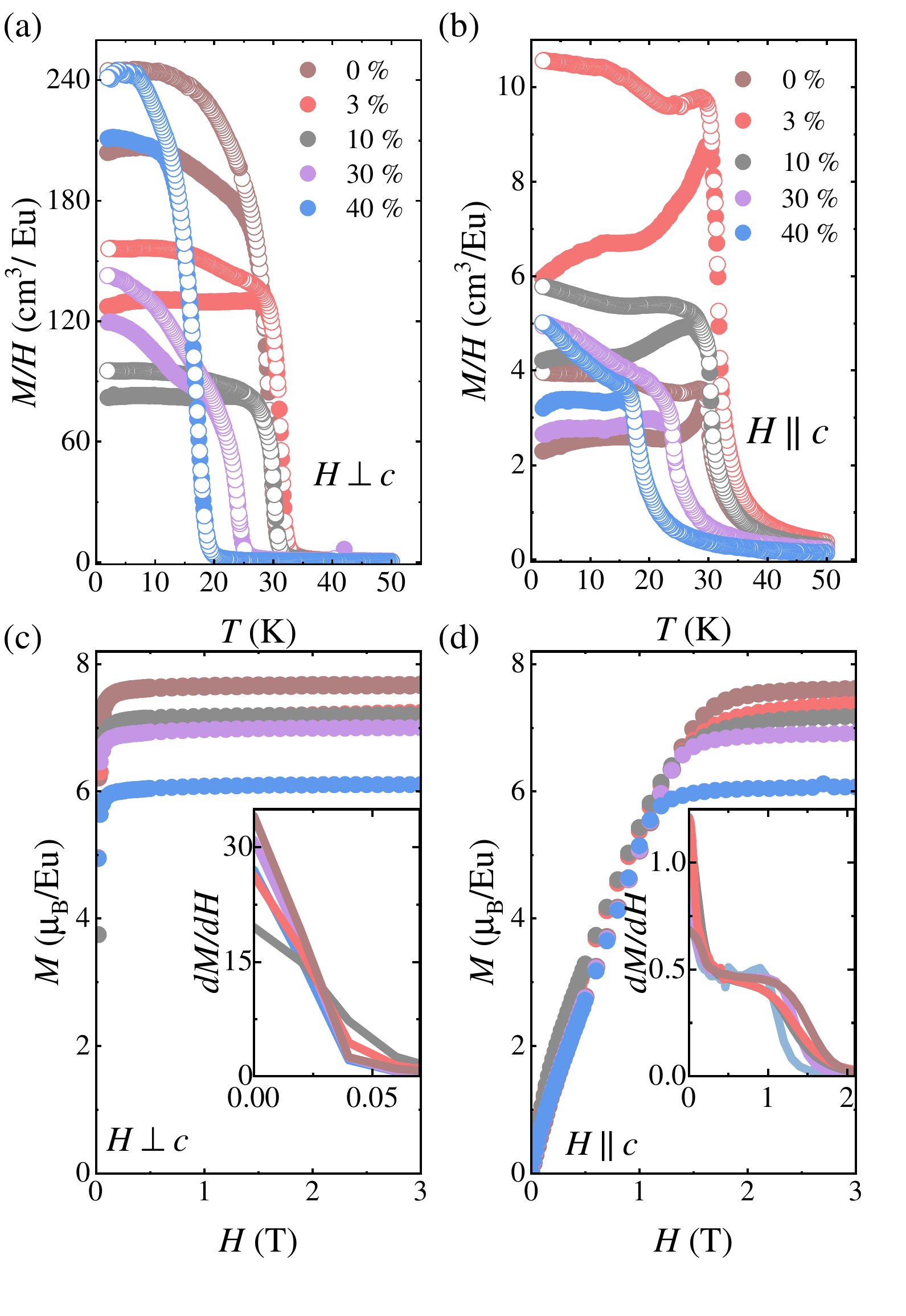}
	\caption{\label{fig:three}  Temperature dependent magnetization of \EBCA\ compositions for a probe field of 50 Oe applied (a) in the \textit{ab} plane and (b) along the \textit{c} axis. Closed circles are for measurements performed in zero field while open circles are for measurements performed in field cooled warming. Magnetization field sweeps for field applied (c) in plane and (d) along the \textit{c} axis. Insets in (c) and (d) show the field derivative $(dH/dM)$ for the two field orientations.}	
\end{figure}

Having achieved some measure of lattice expansion we would like to study the magnetic properties to look for a change in the magnetic ground state. Shown in Fig.~\ref{fig:three} (a-d) are the temperature and field dependent magnetization curves (normalized to Eu content) of all samples used in this study with the field applied both perpendicular and parallel to \c\ and using both field cooled on warming (FC) and zero-field-cooled(ZFC) procedures. Starting with the temperature dependence, we observe clear evidence of magnetic transitions in all samples, which to start we will label as a N\' eel transition (\Tn) based on prior reports. For the parent compound, we see a \Tn\ of $\sim$ 29 K, with a large directional anisotropy where $H\perp\c$ has a response two orders of magnitude larger at 2 K than the $H||c$ direction. While this anisotropy is consistent with previous reports, the \Tn\ we observe here is significantly higher than what was found in the original reports of \Tn $\sim$ 9 K - we will discuss this discrepancy more later \cite{Rahn2018,Schellenberg2011,Wang2016a,Ma2019}.  

We next turn to the field dependent magnetization measured at 2 K (Fig.~\ref{fig:three}(c) and (d).) As seen, we observe behavior similar to previous reports, with a very rapid saturation for a field applied perpendicular to \textit{c} (occurring at $<$ 0.3 T) and a somewhat slower response for a field applied along \textit{c} (saturating at $\sim$ 1.5 T) \cite{Wang2016a, Rahn2018}. For the saturated moment, we find $\sim$7.65 $\mu_B/$Eu which is higher than previously reported but close to the expected value for an Eu$^{2+}$ ion. 

The significantly higher \Tn\ we see in our \ECA\ compound is quite interesting. In a recent report (ref.~\onlinecite{Jo2020}), it was found that by varying synthesis conditions of \ECA\ the magnetic transition and, possibly, correlations could be tuned. In this work a sample with a similar \Tn\ to ours is reported. There the authors suggest the change in \Tn\ due to a slight ($\sim$ 1-4\%) Eu deficiency. Furthermore, they describe the resulting magnetic state as purely FM rather than the previously reported AFM. 

While, we think it likely that sample quality plays a role in the higher \Tn , we are unable to confirm any such Eu vacancies in our sample via EDS or x-ray diffraction. Rather we find several observations to the contrary such as a larger unit cell volume in our sample and an increase in the per Eu saturated moment. Neither of these is consistent with the expectations of Eu vacancies characterized in ref.~\onlinecite{Jo2020}.  As described, our synthesis procedure was quite deliberate and so we continue under the assumption of full Eu occupancy and attribute the change in \Tn\ to different sample qualities. 

Turning to the magnetization curve below the transition we observe different behavior than previously reported. For the in-plane magnetization we find behavior more typical of FM order where the magnetization increases with decreasing $T$ until reaching a saturation point. In prior studies, a downturn was seen at lower temperatures which is not present here. For $H||c$ we find the previously reported temperature dependence with a sharp cusp followed by a drop and a soft shoulder with further decreasing temperature the later of which has been described as due to low laying crystal field levels \cite{Rahn2018, Jo2020}. 

As a second characterization of the magnetism we look at the divergence of the FC and ZFC magnetizations (Fig.~\ref{fig:three} (a) and (b).)  In ref.~\onlinecite{Rahn2018}, the magnetic order along the two crystallographic directions was classified as either FM or AFM based on the FC/ZFC splitting - with a large splitting ostensibly resulting from alignment of the FM components of different magnetic domains. Rahn \textit{et al.} reported a splitting of $\sim$ 15\% for the in-plane magnetization and a splitting of $<$10\%\ for $H||c$, and argued that this was consistent with FM order in-plane and AFM ordering between planes. In our measurements, we see splittings along both directions which are significantly larger (percent changes $>50$\%) than reported previously (measured with the same probe field) and of comparable size along the two directions. Considering the above arguments and those in ref.~\onlinecite{Jo2020}, we suggest that the magnetic structure of the parent compound studied here is likely purely FM with the moments pointing in-plane. Accordingly, for the rest of this paper we will refer to the magnetic transition temperature as \Tc .  

We now consider the effect of Ba substitution on the magnetic properties. As shown in Fig.~\ref{fig:three} (c) and (d) Ba substitution leaves the magnetization field sweeps nearly unchanged qualitatively with similar saturation fields found for all measured compositions. However, a notable decrease in the saturated moment per Eu is seen with a decrease from 7.67$\mu_B$ to 6.1$\mu_B$ from the parent compound to 40\% Ba. Such a change in the per Eu moment is unexpected as Ba$^{2+}$ and Eu$^{2+}$ contribute the same number of electrons and so naively, while the overall formula unit magnetization should decrease with Ba substitution, the moment per Eu is expected to remain unchanged. We can speculate that such a decrease may be due to increased frustration due to the inclusion of non-magnetic Ba on the triangular lattice, or possibly local changes to the single ion physics and/or bonding environment due to the larger Ba$^{2+}$ ion either of which could indicate a route to novel physics \cite{Jlaiel2011}. However, at this point more work is needed to understand the mechanism behind this reduction.   

Looking next at the temperature dependence, we see that Ba substitution at first leaves \Tc\ unchanged, with \Tc\ in the parent compound and the 3\%\ composition being identical within the temperature steps used here. However, as the substitution increases to 10\% Ba we see a decrease in \Tc\ to 27 K, a suppression which continues upon further doping dropping to 20 and 15 K for the 30\% and 40\% samples respectively (we note that we used the same criterion for determining \Tc\ set in ref.~\onlinecite{Rahn2018}.) Considering that the BaCd$_2$As$_2$ end-member is non-magnetic, this suppression of \Tc\ is not unexpected and is consistent with intuition that substituting a non-magnetic ion on a magnetic site should disrupt the magnetic interactions \cite{Yang2013, Kunioka2018}. 

More interestingly, if we look at the magnitude of the magnetization below the transition along the different directions we see different behavior than what might be expected from the field sweeps. As the Ba concentration increases, at first the magnetization along the \textit{c} axis increases, with the largest increase of 160\%\ seen in the 3\% sample (for the ZFC measurements.) Commensurately, for measurements along the in-plane direction we observe at first a steep decrease in the magnetization which drops by $\sim$40\%\ in the 3\%\ sample. As the Ba concentration increases beyond 3\%\ the response along \textit{c} is seen to decrease, nearly returning to the parent compound's value by 40\% while the in-plane response begins to increase until at 40\%\ we again see it nearly return to the parent compound's value. 

Considering the reduction in the in-plane magnetization and the seemingly coupled increase in the magnetization along the \textit{c} axis we propose that the Ba substitution may drive a slight canting of the Eu moment out of the \textit{ab} plane at intermediate concentrations (i.e. at 3-10\% .) Upon further substitution the canting seems to be reduced returning (or nearly returning) to a purely in-plane structure by 40 \%. As a second check of this, we take a closer look at the field sweeps, plotting their field derivative $dM/dH$ in the insets of Fig.~\ref{fig:three}(c) and (d). Though the $M(H)$ curves look similar, the $dM/dH$ curves reveal trends similar to that seen in the $M(T)$. Here both the 3 and 10\%\ samples have a notably slower initial response to the applied field when it is applied in-plane, and a quicker response for $H||c$ than all other samples. This is consistent with the canting scenario, which, at least naively, should lead to a quicker $c$ polarization saturation while requiring a larger field to saturate in-plane. 

The mechanism for how this may happen is as of yet unknown, though previous experimental and theory work more generally on dilute non-magnetic impurities in triangular lattice systems has shown planar to non-planar transitions as a possibility \cite{Nakajima2008, Maryasin2013,Maryasin2015,Smirnov2017}. In any case, the effects of non-magnetic substitution are known to be complex with numerous perturbations such as local structural distortions due to size variance and disruption of magnetic interactions by non-magnetic sites with all of this occurring on an already frustrated triangular lattice. Further studies are needed to understand the magnetic Hamiltonian which drives this behavior as well as diffraction (either with isotopic samples using neutrons or with resonant x-ray techniques) studies to confirm the proposed canting. However, if correct any such canting should have effects on the topological properties as the band structure is predicted to be sensitive to the moment direction \cite{Jo2020, Krishna2018}  

\subsection{\label{AHE} Anomalous Hall Effect and Negative Magnetoresistivity}

\begin{figure}
	\includegraphics[width=\columnwidth]{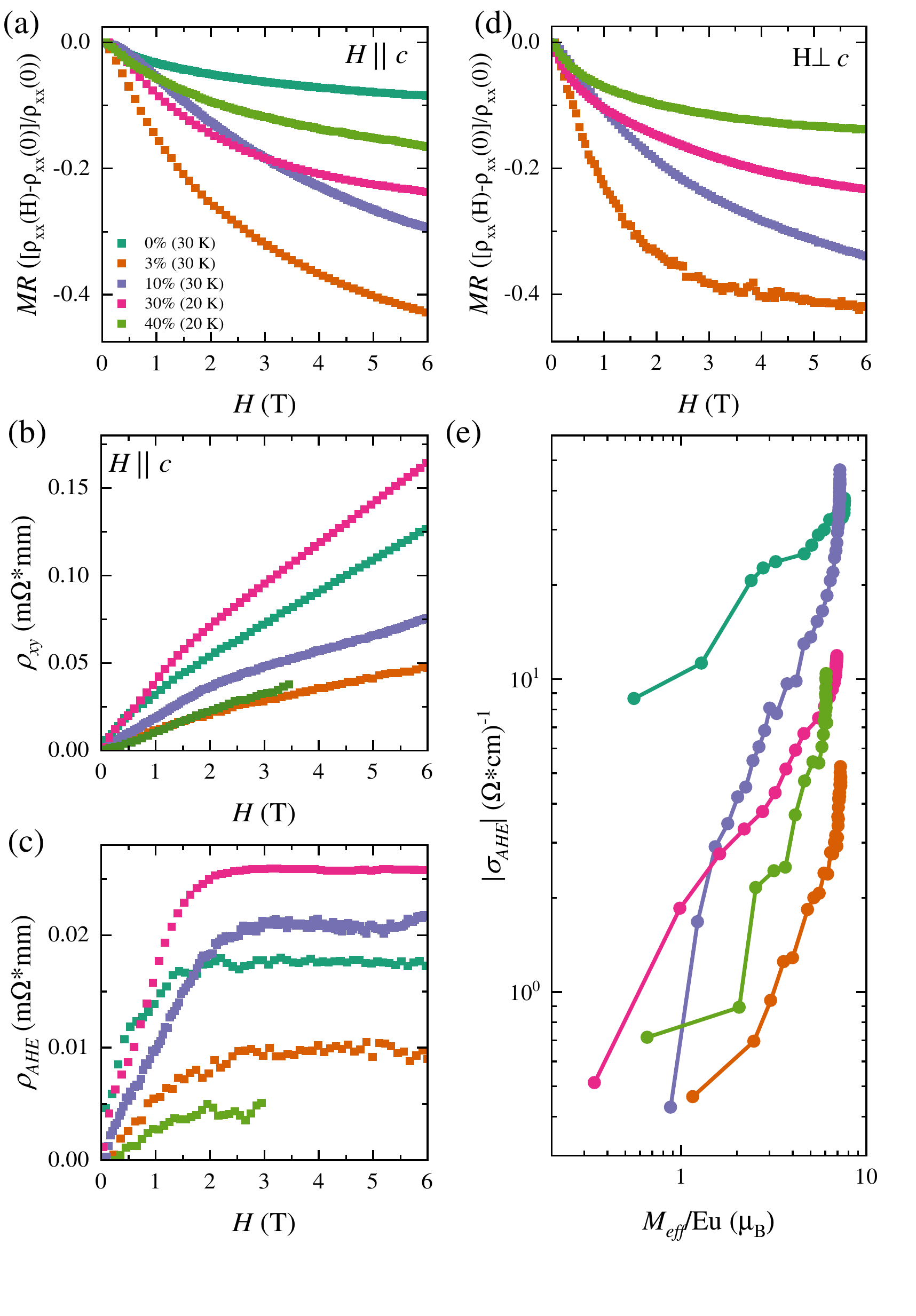}
	\caption{\label{fig:four}  (a)  Magnetoresistance for $H||c$ for the listed compositions at temperatures just above \Tc . (b) Hall resistivity ($\rho_{xy}$) for all samples collected at 2 K. (c) Anomalous Hall resistivity for all samples at 2 K achieved by subtracting the ordinary hall effect contribution from the curves in panel (b). (d) Magnetoresistance of select samples for $H||E$ at the temperatures indicated in panel (a). (d) Absolute value of the anomalous Hall conductivities as a function of magnetization for all compositions at 2 K. The color scheme established in panel (a) for the compositions is continued throughout the figure. }	
\end{figure}

With the magnetic effects of Ba substitution mapped we next look for evidence of changes in the topological physics via the transport properties (Fig.~\ref{fig:four}.) To start we look at the transverse (Hall) resistivity $\rho_{xy}$ (Fig.~\ref{fig:four}(b)) to look for the AHE. For a conducting (or semi-conducting) material a \lq ordinary Hall Effect\rq\ (OHE) is expected which is proportional to the applied magnetic field $H$ - for a FM, an additional AHE is expected  \cite{Nagaosa2010}.  The AHE itself can be derived of numerous phenomena only one of which  (the intrinsic effect due to Berry curvature) is relevant to characterizing the topological properties\cite{Nagaosa2010}. Therefore, we will first separate the AHE and then perform further analysis to identify any intrinsic contribution. 

To extract the AHE we utilize the familiar description of the Hall and AHE in a FM as $\rho_{xy} = R_{0} H_z + R_{s} M_z$ with $R_H$, $B$, $R_s$ and $M_z$ being the OHE coefficient, applied magnetic field perpendicular to the plane of measurement, AHE coefficient and out-of-plane magnetization respectively\cite{Maranzana1967}. To isolate the AHE contribution we perform a linear fit to the high field region of Fig.~\ref{fig:four}(b), obtain $R_0$ and then subtract the $R_0 H_z$ term from $\rho_{xy}$ to get just the anomalous term $\rho_{AHE} = \rho_{xy} - R_0 H_z$ \cite{Meng2019}. We plot the resulting $\rho_{AHE}$ in Fig.~\ref{fig:four}(c). As seen, the $\rho_{AHE}$ contribution for all samples grows for low applied fields and saturates at about 1.5 T - similar to the applied fields at which the magnetization was seen to saturate in Fig.~\ref{fig:three}(d) - indicating the expected $M$ dependence to this AHE term. 

For $\rho_{AHE}$ there are still three possible contributions; skew-scattering and side-jump which are due to disorder and impurity scattering and are dubbed extrinsic effects, and an intrinsic contribution which results from purely quantum effects related to the Berry curvature \cite{Nagaosa2010}. To start isolating the intrinsic contribution, we further separate $\rho_{AHE}$ into two components based on its  dependence on the longitudinal resistivity: $\rho_{AHE} = a\rho_{xx}+b\rho_{xx}^2$. In this form the linear term corresponds to skew-scattering and the quadratic term corresponds to both the side-jump and intrinsic contributions \cite{Zeng2006, Nagaosa2010,Smit1958,Berger1970}. To check for the skew-scattering contribution we use the temperature dependence of $\rho_{xx}$ and the technique described in ref.~\onlinecite{Zeng2006}, where $\rho_{AHE}/(M\rho_{xx})$ is plotted against $\rho_{xx}$ and the linear component is attributed to skew-scattering processes. If a linear regime can be found, it can be fit and subtracted from $\rho_{AHE}$. Doing so here, reveals only a small  skew-scattering contribution to $\rho_{xy}$ (see SM) leaving either the side-jump or the intrinsic effect as dominant \cite{SM}. 

Due to their similar dependence on many of the available tuning parameters, the signatures of the side-jump and intrinsic mechanisms are difficult to separate. In refs~\onlinecite{Manyala2004} and \onlinecite{Jungwirth2002} it is shown that for semiconducting FM plotting the AHE conductivity $(\sigma_{AHE} \sim \rho_{AHE}/\rho_{xx}^2)$ as a function of the magnetization $(M)$ allows for the identification of the intrinsic effect by looking for deviations from $\sigma_{AHE} \propto M$ with such deviations being a sign of strong spin-orbit coupling and therefore a dominant intrinsic term. 

In Fig.~\ref{fig:four}(e) we plot $|\sigma_{AHE}(M)|$ for all samples on a $log - log$ scale by combining the $M(H)$, $\rho_{AHE}(H)$ and $\rho_{xx}(H)$ data (as shown in Figs.~\ref{fig:three}(d), \ref{fig:four}(a-c) and the SM) \cite{SM}. Here initially, $|\sigma_{AHE}|$ increases nearly linearly with $M$ until the effective Eu moment is near its saturation value. As the magnetization increases further $\sigma_{AHE}$ is seen to steeply increase nearly exponentially.  In the theory work of Jungwirth \textit{et al.} such an $M$ dependence was found only for cases of very strong spin-orbit coupling (with a spin-orbit induced gap of $\sim 0.1$ eV) such as is consistent with the heavy Cd ion \cite{Pedrotti1962, Jungwirth2002}. In such cases, their theory predicted a steep increase in $\sigma_{AHE}$ at larger magnetization values similar to that seen here \cite{Jungwirth2002}. We note that we also determined $\sigma_{xy}(M)$ using the procedure used in ref.~\onlinecite{Fang2003} (where temperature dependent $\sigma_{xy}(T)$ and $M(T)$ are used to plot $\sigma_{xy}(M)$ rather than $\sigma_{AHE}(H)$ and $M(H)$ and found similar non-linearities - the results in Fig.~\ref{fig:four}(e) are not due to incomplete subtraction of the $H$ dependence.  As such we take the shape of our $|\sigma_{AHE}(M)|$ to indicate an AHE mainly derived of the intrinsic contribution. We refrain from commenting on the dependence of the AHE on Ba concentration due to the small differences and relative difficulty of the measurements, and instead leave the current analysis as showing that the large intrinsic AHE contribution persists across all compositions, and at least does not show evidence of either a dramatic suppression or enhancement with Ba doping.    


To help study the transport properties across samples, we turn to the longitudinal resistivity as expressed as a \%\ magnetoresistance ($MR = [\rho_{xx}(H)-\rho_{xx}(0)]/\rho_{xx}(0)$) which removes concerns of variations in absolute values which may be affected by systematic errors and derives of a single, large-signal measurement. Fig.~\ref{fig:four}(a) shows the MR of all compositions just above \Tc\ (where the MR was the largest though the trends hold for all temperatures measured) for the $H||c$ $(i.e. H\perp E)$ configuration. All compositions show a significant nMR.  While this is expected for a magnetic material, here we also observe a clear non-monotonic trend in the magnitude of the nMR, with Ba concentration which is not expected \cite{Reed1964}. The nMR increases from the parent compound to 3\%\ and then decreases as the Ba concentration is increased further. Interestingly, though the nMR decreases beyond 3\%\ the magnitude always remains larger than that of the parent compound. This behavior closely mirrors the non-monotonic trend in the magnetization in the $H||c$ channel (Fig.~\ref{fig:three}.)

In addition to the $H\perp E$ configuration, we also measured the MR of several samples with $H||E$ (Fig.~\ref{fig:four}(d).) For a Weyl material, a nMR in this configuration is generally believed due to the chiral anomaly \cite{Son2013,Burkov2015,Ishizuka2019,Li2015b,Li2016}. As for $H\perp E$ all measured samples show nMR of similar magnitude to the perpendicular direction. However, we caution that most treatments of the nMR due to the chiral anomaly are done for non-magnetic materials. Here in the FM \EBCA , it is possible that the nMR in this channel results from the same non-topological mechanism that contributes to the perpendicular direction - the alignment of the PM Eu moments. In which case, attributing this signal to the chiral anomaly becomes more suspect and so we resist drawing any significant conclusions from this observation.    

In general from these MR measurements we see some evidence of a novel effect on the transport properties of the Ba substitution. While the observation of the nMR might not be a smoking gun of Weyl physics in this material, the nonmonotonic behavior that mirrors the magnetization indicates that the doping is having a non-trivial effect. Though one might expect a decrease in the nMR with Ba substitution due to a reduction in the content of the magnetic Eu ion, the observation of a steep increase in the nMR after only a 3\%\ substitution is not expected. Furthermore, though we do observe a reduction upon further substitution, the original value is never recovered, leaving the 40\%\ with a larger nMR than the parent compound. Such results may derive of an additional contribution to the MR not seen in trivial FM and indicate some tuning of the magnetic and/or topological properties which are ostensibly responsible for the nMR in either channel. However, detailed theory work would be needed to extract the various possible contributions and so we leave the discussion here as only an interesting observation. 


\subsection{\label{DFT}Density Functional Theory}

Although a significant amount of theoretical work has been done on this system, both for pure EuCd$_2$As$_2$ and for Ba substitution, we revisit this work with an attempt to answer this operative question: what is the magnetic ground state of the EuCd$_2$As$_2$ system?  While previous work (ref.~\onlinecite{Wang2019}) attempted to answer this question, it used a pseudopotential approach.  Rare earth magnetic properties are notoriously difficult to capture with such approaches, as may be gathered from recent work on potential rare-earth permanent magnet materials (e.g. refs.~\onlinecite{Korner2016} and \onlinecite{Krugel2019}), and so we here revisit this question using a more accurate all-electron approach. 

In the straight GGA we find that a FM configuration of the Eu, both within layers and between layers, is more stable than an AFM layer configuration (termed AFMA in the previous work of ref.~\onlinecite{Wang2019}) by some 3.7 meV per Eu.  This is a rather large value and, although we have not calculated the Curie point, might suggest an ordering point significantly above that observed. With the application of a U of 5 eV and spin-orbit the FM state remains more stable by 0.4 meV per Eu, in contrast to the previous theoretical result where an AFM inter-layer state was asserted. We believe the difference is due to the combination of our use of an all-electron approach, as well as the computational sensitivity of this system. For example, in these calculations a straight GGA calculation of a 4-layer state Eu$_2$Ba$_2$Cd$_8$As$_8$ with the Eu ferromagnetically coupled, resulted in a clearly metastable configuration with one Eu showing the typical 7 $\mu_B$ spin 
moment and the other showing {\it no} moment. These concerns aside, this theoretical finding of an FM ground state is clearly supported by our experimental work so that for this study we regard the matter as resolved.

We have also endeavored to calculate the magnetic anisotropy (MAE), to determine whether the substantial Eu moments prefer a uniaxial or planar orientation, which bears on the Weyl physics. In general, this is a difficult calculation, given that the approximate 1.5 T experimental anisotropy field and $\sim$ 0.6 T saturation magnetization together imply an MAE of no more than 0.2 meV per Eu atom. While the calculations are of sufficient precision and accuracy to resolve this difference, rare earth magnetic anisotropy, even within this accurate all-electron approach, remains a challenging task (please see refs. \onlinecite{Pandey2020, Pandey2018b,Pandey2018b} for examples of reasonably successful efforts in this area.)

Within the computational approach outlined above, and exceedingly careful convergence checks involving, respectively, 5,000, 10,000, 20,000, 40,000 and as many as 75,000 $k$-points, we find a very small, likely planar magnetic anisotropy of order 20 $\mu eV$/Eu, very near the accuracy limit of the calculations. Even at the finest k-point grid this result is not sufficiently converged to usual standards of accuracy (10 \% or less variation).  We can state with confidence, however, that the actual theoretical value is much smaller than the experimental value, and the use of a larger U of 8 eV does not resolve this discrepancy \cite{Ghosh2004}.

From a theoretical perspective, one would expect only a small magnetic anisotropy given than divalent Eu has a nearly exactly half-filled 4$f$ shell, so that Hund's rules would nominally imply zero orbital moment. However, given the non-cubic crystal field there is in the calculations a very small orbital moment of order 0.015 $\mu_B$/Eu, which does lead to a finite, though  small anisotropy. This orbital moment and anisotropy are much smaller than commonly found in rare-earth magnetic materials  and derives principally from the unique character of divalent, half-4-\textit{f}-shell-filled Eu \cite{Larson2003}. 
 
We have also made investigations into the effects of Ba alloying on the magnetic character, given the previous suggestion of associated Weyl physics in a FM state induced via Ba alloying. For this, we have simulated a 50 \% concentration of Ba within two states - one in which the Ba and Eu reside in separate layers (``segregrated FM case") , and one (``mixing FM case")  in which they reside in the same layer.  We have chosen lattice parameters by increasing them proportionately from our experimental values for the 40\% Ba case, relative to the pure Eu case, and separately relaxed the internal coordinates within each configuration. We have also studied an interlayer ``AFMA" state, in which the Eu and Ba layers alternate and there is an additional alternation of Eu moment orientation between Eu layers 2 c-axis lattice constants apart. For simplicity, these calculations were conducted within the straight GGA, without spin-orbit or a 'U'; inclusion of these effects would likely only reduce the interlayer coupling, as in the previous results.

As in the previous results, we find from theory the ``segregated FM" case to be the most energetically stable - some 68 meV per 10-atom cell below the ``mixing FM" case and about 4 meV below the AFMA case \cite{Wang2019}. Note that our XRD results find no evidence for the segregated FM case but rather find the ``mixing FM" structure to be the magnetic ground state. We believe the reason for this discrepancy is the following: At typical synthesis conditions of 1000 K and above, there is a substantial entropic tendency for the divalent Eu and Ba atoms to mix on a layer (the ``mixing FM" state), given that this is a lower-symmetry state than the ``segregated FM" state. This entropic tendency is $-k_B T_{synth} \ln(2)$ per Eu/Ba atom, or about -120 meV per 10-atom unit cell, and generally more than compensates for the energetic stability of the ``segregated FM" state. This suggests that careful low-temperature synthesis
(ideally at temperatures below 600 K) may be necessary to stabilize a ``segregated FM" state for 50\% Ba alloying.

\section{\label{sec:con} Conclusions}

Inspired by DFT predictions of a \textit{c}-polarized FM state in \EfBCA , we performed XRD, magnetization and transport measurements on a series of \EBCA\ samples. Using a careful synthesis procedure we were able to grow high quality single crystals of \EBCA\ with $x\leq 0.4$. Our XRD work showed that Ba does indeed dilate the lattice, and that by $x=0.4$ we achieve a \textit{c} lattice previously predicted to stabilize the \textit{c} polarized FM state. However, magnetization measurements revealed that while Ba substitution does appear to cause out-of-plane canting of the Eu moments, the effect is small and occurs for much lower concentrations than predicted (i.e. 3-10\%.) Additionally, we found that using our deliberate synthesis techniques, all compositions (including the parent compound) show purely FM order with \Tc\ starting at $\sim$ 30 K and being suppressed with increasing Ba concentration. Looking at the transport properties, we found an AHE in all compositions that likely results from a dominant intrinsic contribution. Furthermore, measurements of the magnetoresistance revealed a large nMR in all samples and showed non-monotonic behavior across the series that correlates to the magnetization with the 3 and 10\% samples having the largest nMR. 

More speculatively, we observe a suppression of the saturated moment with Ba substitution likely indicating frustration physics, and see behaviors in the MR in the $H||E$ channel which may hint at the chiral anomaly. Our work shows that Ba substitution has an ability to tune the magnetic and likely topological properties, encouraging follow-up scattering, DFT and synthesis work to determine or confirm the canting, attempt to maximize it via a smaller concentration grid at low substitutions, follow the topological properties as a function of Ba concentration and to study how the numerous perturbations of Ba substitution may tune the magnetic Hamiltonian. This work could lead to a better understanding of how to stabilize the much sought after \textit{c}-FM phase in \ECA\ and in doing so optimize conditions for the realization of Weyl physics in this rare magnetic Weyl semimetal.

\begin{acknowledgments}

The research is partly supported by the US DOE, BES, Materials Science and Engineering Division. The part of the research conducted at ORNL’s High Flux Isotope Reactor was sponsored by the Scientific User Facilities Division, Office of Basic Energy Sciences (BES), US Department of
Energy (DOE). The X-ray diffraction analysis by RC was supported by the US Department of Energy, Office of Science, Basic Energy Sciences, Chemical Sciences, Geosciences, and Biosciences Division. ORNL is managed by UT-Battelle, LLC under Contract No. DE-AC05-00OR22725 for the US Department
of Energy. 

\end{acknowledgments}

\end{document}


\preprint{APS/123-QED}

\title{Supplemental Material for: Evidence of Ba substitution induced spin-canting in the magnetic Weyl semimetal EuCd$_2$As$_2$}

\author{L.D. Sanjeewa}
\thanks{These authors contributed equally}
\affiliation{Materials Science and Technology Division, Oak Ridge National Laboratory, Oak Ridge, TN 37831}
\author{J. Xing}
\thanks{These authors contributed equally}
\affiliation{Materials Science and Technology Division, Oak Ridge National Laboratory, Oak Ridge, TN 37831}
\author{K.M. Taddei}
\thanks{These authors contributed equally}
\affiliation{Neutron Scattering Division, Oak Ridge National Laboratory, Oak Ridge, TN 37831}
\email[corresponding author]{taddeikm@ornl.gov}
\author{D. Parker}
\affiliation{Materials Science and Technology Division, Oak Ridge National Laboratory, Oak Ridge, TN 37831}
\author{R. Custelcean}
\affiliation{Chemical Sciences Division, Oak Ridge National Laboratory, Oak Ridge, TN 37831}
\author{C. dela Cruz}
\affiliation{Neutron Scattering Division, Oak Ridge National Laboratory, Oak Ridge, TN 37831}
\author{A.S. Sefat}
\affiliation{Materials Science and Technology Division, Oak Ridge National Laboratory, Oak Ridge, TN 37831}

\date{\today}

\maketitle

\section{\label{sec:Dstruct}Doping Dependence of Structural Parameters}

Here we list the lattice parameters and internal parameters obtained from our Reitveld refinements of the single crystal x-ray diffraction data shown in Fig. 2 of the main text. 

\begin{table*}[h]
	\caption{\label{tab:one}Crystallographic parameters \EBCA . Parameters determined from Rietveld refinements performed using the 300 K data. The atomic displacement parameters are reported in units of \Asq . }
	\begin{ruledtabular}
		\begin{tabular}{ldddd}
		\multicolumn{1}{l}{} & \multicolumn{1}{c}{\ECA} & \multicolumn{1}{c}{\EnsBCA} & \multicolumn{1}{c}{\EnBCA} & \multicolumn{1}{c}{\EvBCA}   \\
		\hline
\multicolumn{1}{l}{Space Group} & \multicolumn{1}{c}{\Ptmo} & \multicolumn{1}{c}{\Ptmo} & \multicolumn{1}{c}{\Ptmo} & \multicolumn{1}{c}{\Ptmo}   \\
	\multicolumn{1}{l}{$R_{w2}$} & \multicolumn{1}{c}{0.0290} & \multicolumn{1}{c}{0.0522} & \multicolumn{1}{c}{0.067} & \multicolumn{1}{c}{0.0537}  \\
	\multicolumn{1}{l}{$a$ (\AA)} & \multicolumn{1}{c}{4.4412(2)} & \multicolumn{1}{c}{4.4444(3)} & \multicolumn{1}{c}{4.4498(1)} & \multicolumn{1}{c}{4.4750(2)} \\
	\multicolumn{1}{l}{$c$ (\AA)} & \multicolumn{1}{c}{7.3255(8)} & \multicolumn{1}{c}{7.3578(7)} & \multicolumn{1}{c}{7.3702(3)} & \multicolumn{1}{c}{7.5106(3)} \\
		\multicolumn{1}{l}{$V$(\AA$^3$)} & \multicolumn{1}{c}{125.13(2)} & \multicolumn{1}{c}{125.86(3)} & \multicolumn{1}{c}{126.384(8)} & \multicolumn{1}{c}{130.237(5)} \\
	\hline
\multicolumn{1}{l}{Eu ($1a$)}	&		&		&		&		\\
\multicolumn{1}{r}{$U$}	&	0.0159(5)	& 0.0263(5)	 & 0.0267(7)	 &	0.0280(6)	\\
\multicolumn{1}{l}{Ba ($1a$)}	&		&		&		&		\\
\multicolumn{1}{r}{$U$}	&	0.0159(5)	& 0.0263(5)	 & 0.0267(7) &	0.0280(6)	\\
\multicolumn{1}{l}{Cd ($2d$)}	&		&		&		&		\\
\multicolumn{1}{r}{$z$}	& 0.6330(1)	& 0.6330(2)  & 0.6326(2)  &  0.6309(2) 	\\
\multicolumn{1}{r}{$U$}	& 0.017(1)	& 0.0265(5)   & 0.0273(7)  &  0.0282(6)		\\
\multicolumn{1}{l}{As ($2d$)}	&		& 		& 		& 		 		\\
\multicolumn{1}{r}{$z$}	&	0.2469(1)	& 0.2478(2)	 & 0.2480(3) &	0.2545(3)	\\
\multicolumn{1}{r}{$U$}	&	0.014(1)	& 0.035(6)	 & 0.0249(7)	 &	0.0255(6)	\\
\hline
\multicolumn{1}{l}{(Eu-As)}	& 3.1377(6)	& 3.148(1)	& 3.157(1)	&  3.214(1)  	\\
\multicolumn{1}{l}{(Cd-As)} & & & & \\
\multicolumn{1}{r}{basal}	& 2.7110(4)	& 2.7114(7)	& 2.7134(8)	&	2.7231(8)	\\
\multicolumn{1}{r}{apical}	& 2.829(1)	& 2.833(2)	& 2.827(3)	&	2.827(3)	\\
\multicolumn{1}{l}{(Eu-As-Eu)} & 90.1(1)  &  89.79(3) & 89.61(4)  & 88.24(4)   \\

		\end{tabular}
	\end{ruledtabular}
\end{table*}

\newpage

\section{\label{sec:Neutrons}Neutron Diffraction}

A long standing yet vitally important question in \ECA\ is what magnetic structure is actually realized. Though the initial magnetization and transport measurements suggested the desired A-type AFM order this was latter called into question by sophisticated resonant x-ray scattering experiments and modeling \cite{Wang2006a, Rahn2018}. This has been further complicated by recent synthesis studies showing yet another potential magnetic structure to \ECA\ (and even our work here which does the same) \cite{Jo2020}. 
\begin{figure}[h]
	\includegraphics[width=0.7\textwidth]{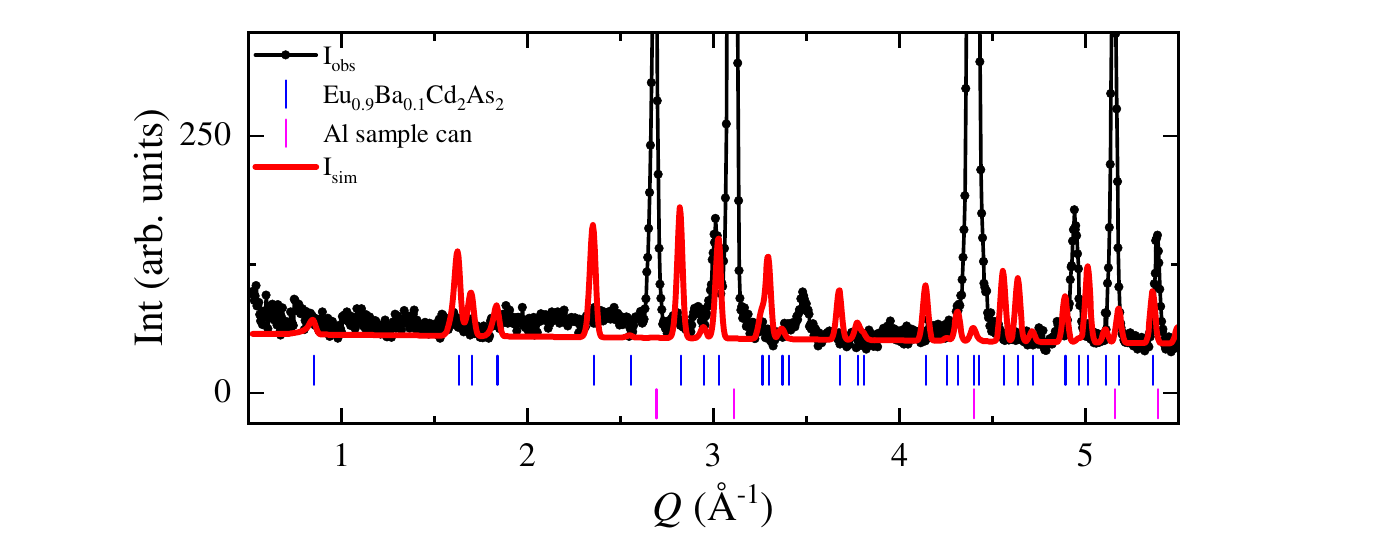}
	\caption{\label{fig:npd} Neutron diffraction pattern of \EnBCA\ collected on HB-2A using a monochromatic incident beam of wavelength $\lambda = 1.12$ \AA . Pattern was collected at 100 K, the indicies for \EnBCA\ and the aluminum sample can are shown as blue and magenta tick marks respectively. For comparison, a simulated powder pattern for \EnBCA\ is shown in red.}	
\end{figure}

This difficulty arises inherently from the composition of \ECA . While normally the magnetic structure is relatively easy to solve using neutron diffraction techniques, here the considerable neutron absorbing cross-sections of both Eu and Cd contraindicate neutron scattering studies. To test this, we performed powder neutron diffraction measurements at the HB-2A powder diffractometer of the HFIR. To prepare the sample, $\sim 100$ mg of single crystals of the \EnBCA\ composition were ground. The resulting powder was then placed on a piece of aluminum foil and arranged in a line of single grains to minimize the neutron path length through the material. The aluminum foil was then rolled and pressed to hold the grains in place and then loaded in a standard Al sample can filled with He. 

In order to minimize absorption, the 117 reflection of the Ge monochromator ($\lambda = 1.12$\AA ) was used giving the highest energy incident beam possible on HB-2A (while still maintaining a useful neutron flux). To maximize flux we used a collimator configuration of op-op-12' for the pre-monochromator, pre-sample and pre-detector collimators respectively \cite{Calder2018}. We also performed measurements using the 1.54 \AA\ incident wavelength due to its higher flux however, these are not shown here as there was no improvement in the collected pattern. 

The resulting neutron powder diffraction pattern collected at 100 K is shown in fig.~\ref{fig:npd} together with a simulated \EnBCA\ diffraction pattern. Despite our efforts, the obtained pattern is not suitable for either qualitative or quantitative analysis - with only a couple peaks visible from the main phase and an obvious non-isotopic absorption as a function of $Q$ presumably due to different neutron path lengths and anisotropy in the grain shapes. Notably, several strong peaks predicted for low $Q$ have no intensity in our pattern. As this experiment was optimized to minimize absorption (our ostensible greatest neutron path-length being the diameter of a single grain of magnitude $<$ 0.5mm) we suggest these results contraindicate neutron scattering in these compounds in non-isotopic samples short of using sub-millimeter sized single crystals sanded into a cylindrical symmetry, or a hot neutron source.

\section{\label{sec:res}Transport}

\begin{figure}[h]
	\includegraphics[width=\textwidth]{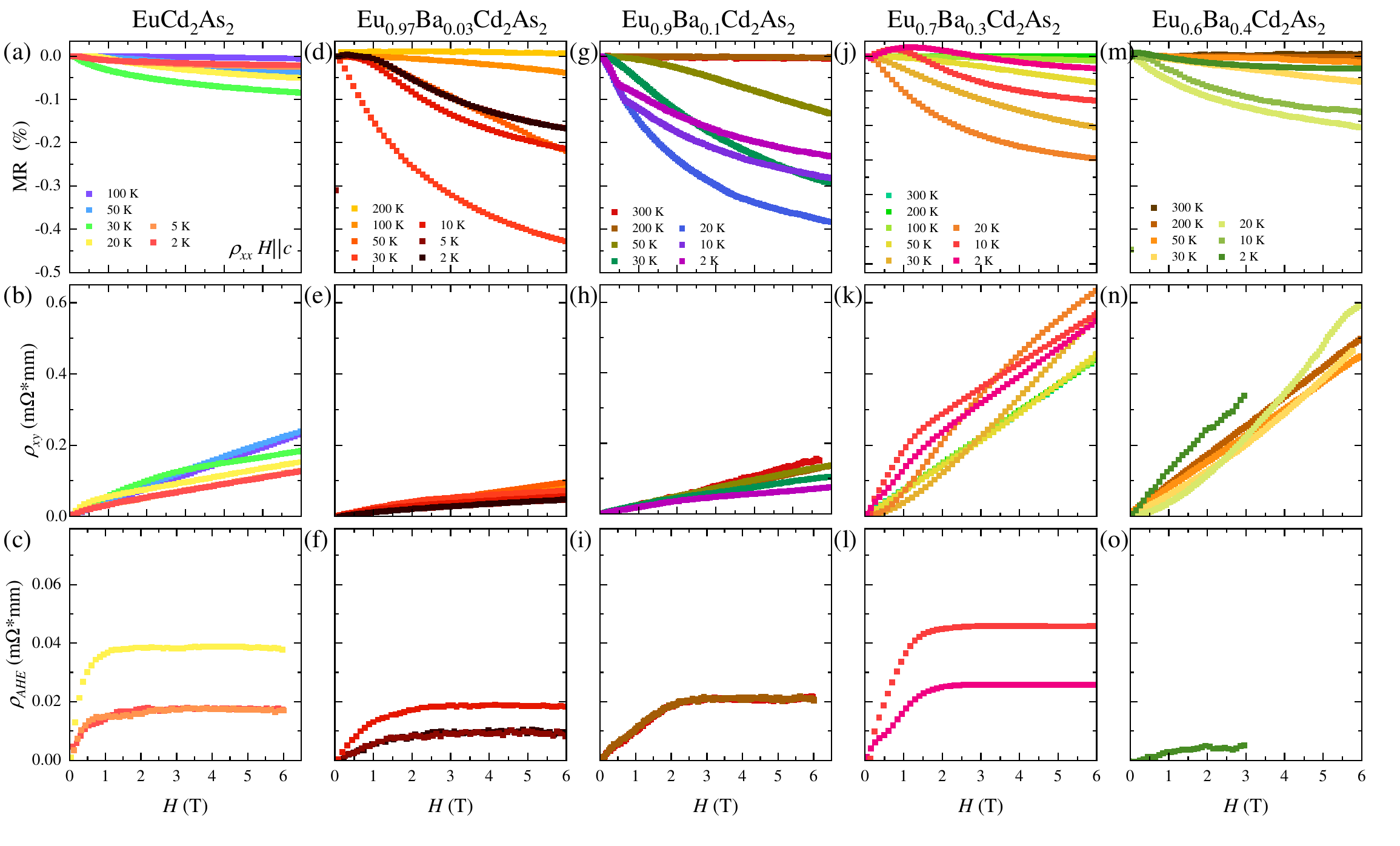}
	\caption{\label{fig:AHE} Transport measurements showing the longitudinal magnetoresistance (with $H||c$), Hall Effect and isolated Anomalous Hall Effect for (a-c) \ECA , (d-f) \EnsBCA , (g-i) \EnBCA , (j-l) \EsBCA\ and (m-o) \EvBCA . }	
\end{figure}

In Fig.~\ref{fig:AHE} we report the resistivity measurements for all samples at every measured temperature. While we mainly include these for completeness, we would also like to point out two additional observations here not discussed in the main text. First that in the MR, along with the general nMR, there is a second feature (most clearly visible above \Tc\ in \EsBCA ) of a crossover from initial positive MR to negative MR which has also been suggested as indicative of topological physics. Secondly, in the plots of $\rho_{AHE}$ we note that the AHE is largest not at base temperature but just below \Tc . We see this effect in all samples with the appropriate temperatures but have no explanation for it at the moment outside of some hand-waving supposition that fluctuations may still be present just below \Tc .  The effect is strongly dependent on temperature and is seen only just below \Tc . Comparisons, for instance, of 2 and 5 K show no difference. 

We note that due to the crystal size, morphology and (potentially) FM domain formation measurements of the HE were challenging - for the \EvBCA\ sample not all temperatures and field ranges are present due to these difficulties which were more pronounced in the samples with higher Ba concentrations.   

\begin{figure}[h]
	\includegraphics[width=0.6\textwidth]{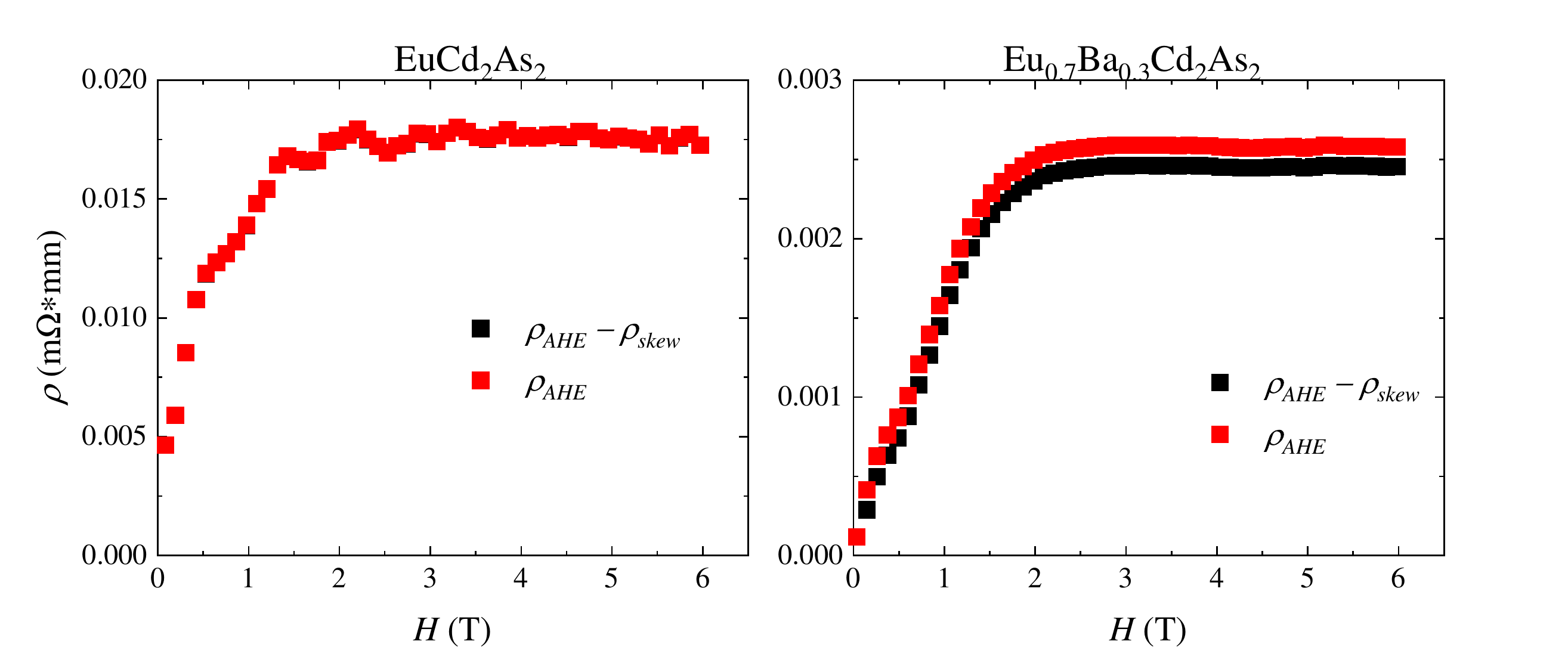}
	\caption{\label{fig:skew} Comparison of the total $\rho_{AHE}$ to $\rho_{AHE} - \rho_{skew}$ for \ECA\ and \EsBCA\. }	
\end{figure}

As mentioned in the main text, attempts where made to remove the skew scattering contribution to the AHE by fitting the linear region (if one exists) of $\rho_{AHE}/(M\rho_{xx})$ and then using the slope to define a linear term which derives of skew scattering which can then be subtracted from the AHE (see ref.~\onlinecite{Zeng2006} for details.) The resulting comparison of the full AHE and the combined side-jump/instrinsic components are shown in Fig.~\ref{fig:skew} for \ECA\ and \EsBCA . As shown this leads to only a small correction to $\rho_{AHE}$ and so we conclude that skew-scattering in this material makes only a minimal contribution to the AHE.   

%